\documentclass[amsmath,amssymb,superscriptaddress,twocolumn]{revtex4-1}

\usepackage{graphicx}
\usepackage{dcolumn}
\usepackage{bm}
\usepackage{hyperref}
\usepackage{stmaryrd}
\usepackage{color}
\usepackage{amsmath}

\DeclareMathAlphabet\mathbfcal{OMS}{cmsy}{b}{n}
\def\vec#1{{{\underline{#1}}}}
\def\tensor#1{{\underline{\underline {#1}}}}

\def\d {{\rm d}}

\def\nm {{\rm nm}}
\def\Tr {{\rm Tr\,}}

\begin{document}

\title{Effects of disorder and chain stiffening on the elasticity of flexible polymer networks}

\author{Christiane Caroli}
\affiliation{INSP, Universit\'e Pierre et Marie Curie-Paris 6, CNRS, UMR 7588, 4 place Jussieu, 75252 Paris Cedex 05, France}
\author{Ana\"el Lema\^{\i}tre}
\affiliation{NAVIER, UMR 8205, \'Ecole des Ponts, IFSTTAR, CNRS, UPE, 2 all\'e K\'epler, F-77420 Marne-la-Valle, France}

\date{\today}

\begin{abstract}
We examine how the distribution of contour lengths and the high-stretch stiffening of individual chain segments affect the macroscopic shear modulus of flexible polymer gels, using a 2D numerical model, in which polymer segments form a triangular network and disorder is introduced by varying their contour lengths. We show that in the relevant parameter range: (i) the non-affine contribution to the shear modulus is negligible, i.e. the Born approximation is satisfactory; (ii) the shear modulus is dominated by the contribution originating from equilibrium chain tensions. Moreover, mechanical equilibration at the nodes induces specific correlations between the end-to-end distances and contour lengths of chain segments, which must be properly accounted for to construct reasonable estimates of chain pressure and shear modulus.
\end{abstract}

\maketitle

\section{Introduction}

In the last decades, a lot of attention has been dedicated to understanding how the macroscopic elasticity of polymer hydrogels is controlled by the properties of single polymer strands and by network disorder. Most of these studies~\cite{Kroy2006,BroederszMacKintosh2014,PritchardHuangTerentjev2014} have focussed on the case of filamentous networks, in which the constitutive polymers have persistence lengths $\ell_p$ larger that the network mesh size $\xi$ -- which is relevant to many biopolymers. They have shown, in particular, that large chain stiffnesses give rise to spectacular mechanical responses, such as pre-stress-induced stiffening or softening, negative normal stress, etc. Few works~\cite{CarrilloMacKintoshDobrynin2013,MengTerentjev2016}, however, have addressed recently the case -- which is relevant to many hydrogels, such as gelatin or synthetic polymer gels -- where the persistence length $\ell_p$ of polymer strands lies in an intermediate range between the monomer size $a$ and the mesh size $\xi$.

The elasticity of such ``flexible polymer networks'' is usually treated in the framework of the classical theory of rubber elasticity, which suggests that their shear modulus is $\sim k_BT/\xi^3$ (in 3D). This expression is often used in practice to estimate network mesh-sizes from measurements of the shear modulus, but only arises from approximate scaling arguments~\cite{deGennes1979}, or from a mean-field theory~\cite{Treloar1949} where the end-to-end distances of chain segments are supposed to follow the same Gaussian statistics as if they were independent. This assumption overlooks the fact that chains fluctuate around a state of mechanical equilibrium; as we shall see, it amounts to a very stringent ansatz about the correlations between end-to-end and contour lengths.

In this paper, we study in detail how the distribution of contour lengths and the high-stretch stiffening of individual chain segments affect the macroscopic shear modulus. We address these issues using a 2D numerical model of gel, in which disorder is introduced by varying the contour lengths of polymer strands that form a triangular network. The model takes into account the existence of excluded volume effects \`a la Flory-Rehner~\cite{Onuki2002}, and is tested for different expressions of the single chain response to large extensions. The shear modulus of such a system can be decomposed into a sum of three terms:
\begin{equation}\label{eq:G:decomp}
G=-P^{\rm ch}+C+G^{\rm na}
\end{equation}
which are respectively the opposite of the chain pressure $ P^{\rm ch}$, the elastic ``constant'' $C$, and the non-affine term $G^{\rm na}$. Even though the respective importance of these three terms depends on disorder strength and swelling level as well as on the importance of stretch-stiffening, we show that, in the range of parameter values relevant for usual flexible gels, these three contributions are ordered according to the following hierarchy: $-P^{\rm ch}\gg C\gg G^{\rm na}$. Namely, the non-affine contribution to the shear modulus is small enough for the Born approximation $G\simeq-P^{\rm ch}+C$ to provide, for all practical purposes, a very satisfactory estimate of the shear modulus. The elastic constant $C$ contributes at most a fraction of $G$ less than typically 15--20\% and is all the weaker that the persistence length is small. The chain pressure term is always dominant, but its value depends on both the distribution of chain segment stiffnesses and on the accomodation of elastic disorder by mechanical equilibration. This hierarchy of the contributions is specific to flexible polymer gels, in constrast with semi-flexible or rigid polymer networks.


\section{Model}

We choose to specialize to 2D networks with a fixed, triangular topology, as illustrated on Fig.~\ref{fig:pictures}. The nodes represent permanent crosslinks and the links flexible polymer strands that rotate freely at the nodes. The number of monomers on the strand connecting nodes $i$ and $j$ is denoted $N_{ij}$.

\subsection{System free-energy}
\label{sec:system}
We define the system free-energy as $\mathcal{F}=\mathcal{F}^{\rm ch}+\mathcal{F}^{\rm Fl}$, where the two terms account respectively for:\\
(i) the elastic free-energy of individual strands in an ideal solvent,
\begin{equation}\label{eq:fch}
\mathcal{F}^{\rm ch}=\sum_{\{ij\}} F_{ij}(r_{ij})
\end{equation}
where the sum runs over all strands $\{ij\}$, with $r_{ij}$ the distance between nodes $i$ and $j$, and
the subscript $ij$ in $F_{ij}$ accounts for its dependence on $N_{ij}$.\\
(ii) a Flory-Rehner-like contribution~\cite{Onuki2002} that accounts for excluded volume effects. To model it, we consider that the number of monomers lying in a given network triangle $\{ijk\}$ is $\frac{1}{2}({N_{ij}+N_{jk}+N_{ik}})$. The average monomer concentration in a network triangle is thus
\begin{equation}
c_{ijk}=\frac{1}{2}\frac{N_{ij}+N_{jk}+N_{ik}}{A_{ijk}}
\end{equation}
with $A_{ijk}$ the triangle area. In the spirit of the Flory-Rehner mean-field approximation~\cite{Onuki2002}, we then set:
\begin{equation}\label{eq:Flory}
\mathcal{F}^{\rm Fl}=\frac{k_BT}{2}(1-2\chi)a^2\sum_{\{ijk\}}c_{ijk}^2A_{ijk}\equiv\sum_{\{ijk\}}F_{ijk}(A_{ijk})
\end{equation}
where the sum runs over all triangles, and $a^2(1-2\chi)$ is the 2D excluded volume parameter.

Now we need to specify the expression of the elastic free-energy $F_{ij}$ of individual chains that enters equation~(\ref{eq:fch}). We consider that the persistence length $\ell_p$ is small compared with the average mesh size. So long as the end-to-end distance $r_{ij}$ of any strand is much smaller than its contour length $a N_{ij}$, elasticity is purely entropic, and we can use for $F_{ij}$ the standard Gaussian expression~\cite{RubinsteinColby2003}:
\begin{equation}\label{eq:G}
F_{ij}^{\rm G}(r_{ij})=\frac{k_BT}{2a\ell_p}\,\frac{r_{ij}^2}{N_{ij}}
\end{equation}
It is well-known, however, that this expression is insufficient at high stretch, i.e. when $x_{ij}=r_{ij}/(aN_{ij})$ approaches 1.

A standard model for the large stretch response is the freely jointed chain~\cite{RubinsteinColby2003}, for which the relation between elongation and applied force is provided, in 3D, by a Langevin function. Although the associated free-energy does not possess an explicit analytic expression, a satisfactory approximation is provided by the Cohen expression~\cite{Cohen1991}, which interpolates between the small stretch ($x\ll1$) Gaussian and the nearly taut ($x\to1$) limits. In Appendix~\ref{app:fjc} we derive the analogous expression for a 2D freely-jointed chain containing $aN_{ij}/(2\ell_p)$ Kuhn segments; it reads:
\begin{equation}\label{eq:fj}
F_{ij}^{\rm FJ}(r_{ij})={k_BT}\,\frac{a\,N_{ij}}{2\ell_p}\,w\left(\frac{r_{ij}}{aN_{ij}}\right)
\end{equation}
with
\begin{equation}
w(x)=\frac{1}{2}\left(x^2-\log\left(1-x^2\right)\right)
\end{equation}
As needed, $F^{\rm FJ}$ reduces to the Gaussian expression in the limit $x\ll1$.

It is known that the logarithmic growth of the freely-jointed chain free-energy underestimates the hardening of the single chain response at high stretch levels. This effect certainly becomes all the more important that the persistence length increases and, in the limit of high stiffnesses, the worm-like chain (WLC) model provides a much better description of the stretch-force relation. To tackle the problem of intermediate persistence lengths, Blundell and Terentjev~\cite{BlundellTerentjev2009} have proposed a model of the single chain response that interpolates between the Gaussian and the WLC relations. This provides a third model for the elastic free-energy\footnote{In order to ensure that $F_{ij}^{\rm BT}$ matches $F_{ij}^{\rm G}$ in the $Na/\ell_p\gg1$, $x\ll1$ limit, our definition of the persistence length differs by a factor of $4/\pi$ from Blundell and Terentjev's}:
\begin{equation}
F_{ij}^{\rm BT}(r_{ij})=k_BT\frac{2\pi\ell_p}{aN_{ij}}\left(1-x_{ij}^2\right)
+{k_BT}\frac{aN_{ij}}{2\ell_p}\frac{1}{1-x_{ij}^2}
\end{equation}
with $x_{ij}=r_{ij}/(aN_{ij})$.

In the following, we will probe the linear and non-linear elastic response of networks using these three expressions for the single chain free-energy.

\subsection{Units and parameter ranges}

We will use parameter values in ranges that are reasonable for the thoroughly investigated hydrogels of gelatin, for which the monomer size $a\simeq3\AA$, persistence length $\ell_p\simeq 7a$, and $1-2\chi\simeq2.10^{-2}$~\cite{BohidarJena1994}. Typical mesh sizes $\xi$ lie in the $10\nm$ range, i.e. $\sim30a$. A rough evaluation of the average number of monomers per strand is $N\sim\xi^2/(2a\ell_p)$, on the order of 100.

When presenting numerical data, the monomer size $a$ and $k_BT$ will be taken as units of length and energy.

\subsection{Network disorder}

\label{sec:disorder}

To construct a triangular network, the node points are initially placed on a Bravais lattice with vectors $(\xi,0)$ and $(\xi/2,\sqrt{3}\,\xi/2)$, in a biperiodic cell of extension $(\xi M_x,\sqrt{3}\,\xi M_y/2)$, with $M_x$ and $M_y$ integers. There are $N^{\rm node}=M_xM_y$ nodes for a cell area $A=\sqrt{3}\xi^2 M_xM_y/2$, i.e. an average areal density of nodes $\rho=2/(\xi^2\sqrt{3})$. The number of strands is $N^{\rm ch}=3N^{\rm node}$ and the number of triangles $N^{\Delta}=2N^{\rm node}$.

\begin{figure}
\includegraphics[width=0.22\textwidth]{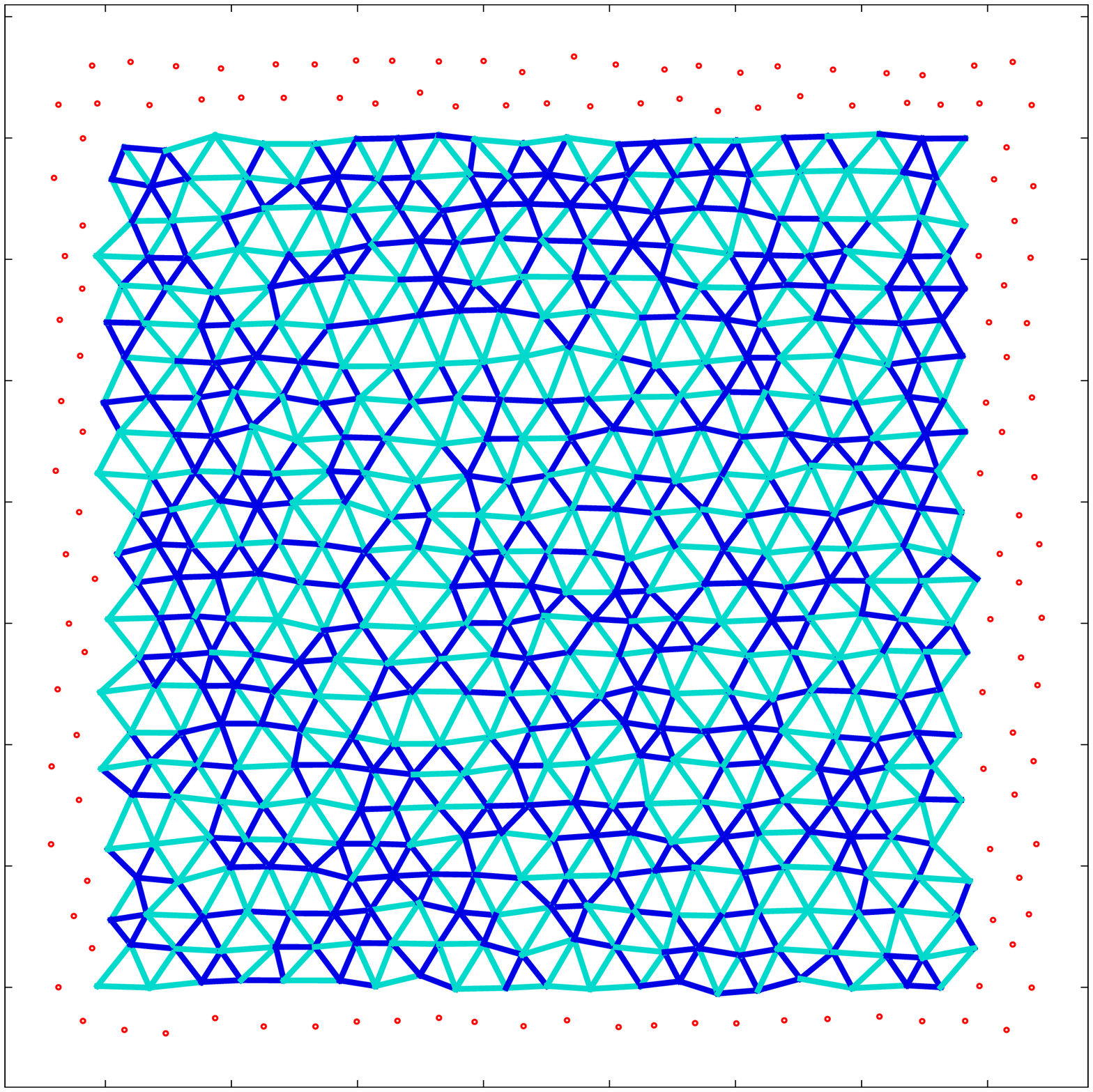}
\includegraphics[width=0.22\textwidth]{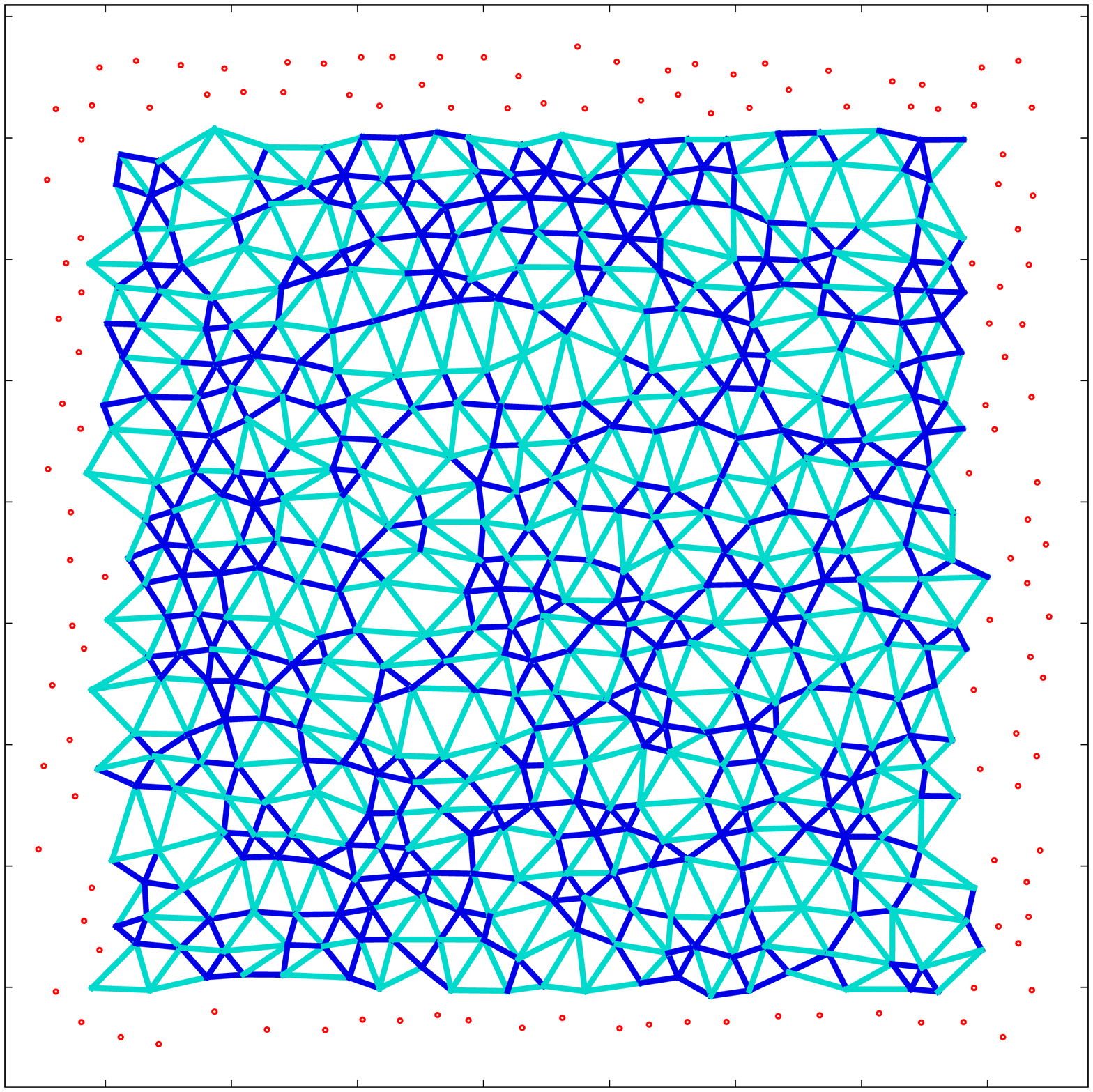}
\caption{\label{fig:pictures}
(Color online) Mechanically equilibrated disordered network configurations for the FJ chain model, using the bimodal distributions $B_{100,30}$ (left) and $B_{100,50}$ (right) of the $N_{ij}$'s, with parameters $\ell_p=7a$, $1-2\chi=2.10^{-2}$, $\xi=30a$. Short and long chains are drawn in dark and light blue.
}
\end{figure}

The network topology being fixed, disorder is introduced via the values of the monomer numbers $N_{ij}$, which we take to be random and uncorrelated variables. For the sake of simplicity, and to facilitate the qualitative analysis of disorder effects, we assume their distribution $B_{N,\Delta}(N_{ij})$ to be bimodal:
\begin{equation}\label{eq:bim}
\begin{aligned}
B_{N,\Delta}(N_{ij})&=\quad\frac{1}{2} \qquad  &&{\rm if} \quad N_{ij}=N-\Delta\\[-2mm]
& && {\rm or } \quad N_{ij} =N+\Delta\\
&= \quad0 &&{\rm otherwise}
\end{aligned}
\end{equation}

For any realization of the set $\{N_{ij}\}$, mechanical equilibrium is then found by minimizing the total free-energy of the system, which results in a distorted network. This is illustrated on Fig.~\ref{fig:pictures}, which displays two mechanically equilibrated configurations with the bimodal disorder defined by $B_{100,30}$ and $B_{100,50}$.

\section{Elastic network response}

\subsection{Affine vs non-affine contributions}
\label{sec:naff}
To compute elastic constants, we rely on the general formalism developed in~\cite{LemaitreMaloney2006} for the elastic response of disordered solids. Its main lines are briefly summarized as follows.

Let us consider some initial (reference) state about which we compute the elastic response. The externally imposed macroscopic deformation about this initial state is specified via the strain tensor $\tensor\Lambda$. In the initial state, $\tensor\Lambda=\tensor1$ and the nodes assume equilibrium positions denoted $\vec r_i(\tensor\Lambda=\tensor1)$. Under deformation, an arbitrary configuration of the system is defined by the macroscopic strain $\tensor\Lambda$ and the node positions $\{\vec r_i\}$. For any $\vec r_i$, we define its zero strain antecedent as $\mathring{\vec r}_i\equiv\tensor\Lambda^{-1}\cdot\vec r_i$. Clearly $\mathring{\vec r}_i\equiv\vec r_i$ in the initial configuration, where $\tensor\Lambda=\tensor1$.

The Born approximation of affine deformation amounts to assuming that node positions vary with $\tensor\Lambda$ as $\vec r_i=\tensor\Lambda\cdot\mathring{\vec r}_i(\tensor1)$ with fixed antecedents that coincide with the initial node positions. To separate the affine and non-affine contributions to the elastic response, it is convenient to write formally the total free-energy $\mathcal{F}\left(\{\vec r_i\},\tensor\Lambda\right)$ of the deformed system in terms of the antecendents $\mathring{\vec r}_i=\tensor\Lambda^{-1}\cdot\vec r_i$ of the (a priori arbitrary) node positions. This is realized by writing
\begin{equation}
\mathcal{F}\left(\{\vec r_i\},\tensor\Lambda\right)
=\mathcal{F}\left(\{\tensor\Lambda\cdot\mathring{\vec r}_i\},\tensor\Lambda\right)
\equiv
\mathring{\mathcal{F}}\left(\{\mathring{\vec r}_i\},\tensor\Lambda\right)
\end{equation}
which defines $\mathring{\mathcal{F}}$.

We are interested in the static elastic response, which entails that, under loading, the system remains at mechanical equilibrium, i.e. that the force on each node vanishes at all times:
\begin{equation}\label{eq:mech}
\vec f_i=-\frac{\partial \mathcal{F}}{\partial\vec r_i}=-\frac{\partial \mathring{\mathcal{F}}}{\partial\mathring{\vec r_i}}\cdot\tensor\Lambda^{-1}=0
\end{equation}
We denote $\vec r_i(\tensor\Lambda)$ the node positions at mechanical equilibrium under strain $\tensor\Lambda$. In a disordered system, as $\tensor\Lambda$ varies, the $\vec r_i(\tensor\Lambda)$ follow trajectories that, in general, are not affine. It means that their antecedents $\mathring{\vec r}_i(\tensor\Lambda)\equiv\tensor\Lambda^{-1}\cdot\vec r_i(\tensor\Lambda)$ are not fixed, but vary with $\tensor\Lambda$. Their trajectories are specified by derivating, with respect to each component $\Lambda_{\alpha\beta}$ of the strain tensor, the condition
\begin{equation}
\frac{\partial \mathring{\mathcal{F}}}{\partial\mathring{\vec r_i}}=0
\end{equation}
which is an immediate consequence of Eq.~(\ref{eq:mech}).
In the limit $\tensor\Lambda\to\tensor1$, it comes:
\begin{equation}\label{eq:naff}
\left.\mathcal{H}_{ij}\cdot\frac{\partial\mathring{\vec r}_j}{\partial\Lambda_{\kappa\chi}}\right|_{\tensor\Lambda\to\tensor1}=\vec\Xi_{i,\kappa\chi}
\end{equation}
(note that we use the convention of implicit summation on repeated indices). Here,
\begin{equation}
\mathcal{H}_{ij}=\left.\frac{\partial^2\mathring{\mathcal{F}}}{\partial\mathring{\vec r}_i\partial\mathring{\vec r}_j}\right|_{\tensor\Lambda\to\tensor1}
\end{equation}
is the Hessian matrix in the reference (initial) state, and the vector field $\vec\Xi_{i,\kappa\chi}$ is defined as:
\begin{equation}\label{eq:Xi}
\vec\Xi_{i,\kappa\chi}=-\left.\frac{\partial^2\mathring{\mathcal{F}}}{\partial\Lambda_{\kappa\chi}\partial\mathring{\vec r}_i}\right|_{\tensor\Lambda\to\tensor1}
\end{equation}
Note that taking partial derivatives of $\mathring{\mathcal{F}}$ with respect to strain components amounts to  varying $\tensor\Lambda$ at constant $\{\mathring{\vec r}_i\}$, i.e. to performing affine deformations about state $\{\mathring{\vec r}_i\}$. Since moreover, in the limit $\tensor\Lambda\to\tensor 1$, $\{\mathring{\vec r}_i\}$ tends to the initial (reference) configuration, it turns out that $\vec\Xi_{i,\kappa\chi}\delta\Lambda_{\kappa\chi}$ can be interpreted as the force induced by an infinitesimal affine deformation $\delta\Lambda_{\kappa\chi}$~\cite{LemaitreMaloney2006}. In view of equation~(\ref{eq:naff}), the non-affine displacement field characterized by $\left.\frac{\partial\mathring{\vec r}_j}{\partial\Lambda_{\kappa\chi}}\right|_{\tensor\Lambda\to\tensor1}$ can be interpreted as the linear elastic response of the system to this field of virtual forces.

This framework can be used to write explicit expressions for stresses and elastic stiffnesses, which are first and second derivatives of the free-energy with respect to strain. In particular, it has been shown that elastic stiffnesses~\cite{BarronKlein1965}, defined as
\begin{equation}\label{eq:S}
S_{\alpha\beta\kappa\chi}\equiv\frac{1}{A}\,
\left.\frac{\partial^2\mathcal{F}}{\partial\Lambda_{\alpha\beta}\partial\Lambda_{\kappa\chi}}\right|_{\tensor\Lambda\to\tensor1}
\end{equation}
with $A$ the system area, can be decomposed as~\cite{LemaitreMaloney2006}:
\begin{equation}\label{eq:S}
S_{\alpha\beta\kappa\chi}=S_{\alpha\beta\kappa\chi}^{\rm Born}+
S_{\alpha\beta\kappa\chi}^{\rm NA}
\end{equation}
where
\begin{equation}\label{eq:Born}
S_{\alpha\beta\kappa\chi}^{\rm Born}\equiv\frac{1}{A}\,
\left.\frac{\partial^2\mathring{\mathcal{F}}}{\partial\Lambda_{\alpha\beta}\partial\Lambda_{\kappa\chi}}\right|_{\tensor\Lambda\to\tensor1}
\end{equation}
is the Born approximation for the stiffness tensor, which assumes that the nodes follow affine trajectories, and the non-affine contribution
\begin{equation}\label{eq:na}
S_{\alpha\beta\kappa\chi}^{\rm NA}=-\frac{1}{A}\,\vec\Xi_{i,\alpha\beta}\cdot(\mathcal{H}^{-1})_{ij}\cdot\vec\Xi_{j,\kappa\chi}
\end{equation}
results from the non-affinity of the displacement field.

\subsection{Microscopic expression of the stress tensor}

Since $\mathcal{F}=\mathcal{F}^{\rm ch}+\mathcal{F}^{\rm Fl}$, the macroscopic Cauchy stress tensor is a sum of a chain and a Flory contribution
\begin{equation}
\tensor\sigma=\frac{1}{A}\left.\frac{\partial\mathcal{F}}{\partial\tensor\Lambda}\right|_{\tensor\Lambda\to\tensor1}
=\tensor\sigma^{\rm ch}+\tensor\sigma^{\rm Fl}
\end{equation}
The ``chain stress'' is given by the classical expression for systems with pair interactions, namely:
\begin{equation}\label{eq:sigma:ch}
    \tensor\sigma^{\rm ch}
    =\frac{1}{A}\sum_{i<j}\,F_{ij}'(r_{ij})\frac{\vec r_{ij}\,\vec r_{ij}}{r_{ij}}
\end{equation}
with $\vec r_{ij}=\vec r_j-\vec r_i$. The ``Flory stress'', derived in Appendix~\ref{app:details}, which reads
\begin{equation}\label{eq:sigma:Fl}
    \tensor\sigma^{\rm Fl}
    =\frac{1}{A}\sum_{ijk}\,F'_{ijk}(A_{ijk})\,A_{ijk} \,\tensor1
\end{equation}
is diagonal under any state of deformation, as expected from the microscopically isotropic character of the Flory interaction; it hence only contributes to the osmotic pressure
\begin{equation}
P=-\frac{1}{2}\,\Tr\tensor\sigma
\end{equation}
but not to deviatoric stresses.

\subsection{Response to simple shear}

Since hydrogels are incompressible on time scales where poroelasticity is irrelevant, their elasticity is characterized, for all practical purposes, by their response to shear. We thus focus here on the case of simple shear deformation, for which
\begin{equation}
\tensor\Lambda = \
\left(\begin{matrix}
1\, & \gamma \\
0\, & 1
\end{matrix}\right)
\end{equation}
In this case, the linear elastic response is characterized by the shear modulus $G=S_{xyxy}$, which, according to equations~(\ref{eq:S})-(\ref{eq:na}), can be decomposed into a Born and a non-affine contributions.

We first calculate explicitly the Born contribution $G^{\rm Born}$, which is obtained under the assumption of affine node displacement. Denoting $\vec r_{i}^0$ the initial node positions, any strand end-to-end vector $\vec r_{ij}^0$ is transformed into $\vec r_{ij}=\tensor\Lambda\cdot{\vec r}_{ij}^0$, i.e.:
\begin{equation}
\left\{
\begin{split}
x_{ij} &= x_{ij}^0+\gamma\,y_{ij}^0\\
y_{ij} &= y_{ij}^0\\
\end{split}
\right.
\end{equation}
Under such an affine displacement, which preserves areas, the Flory free-energy $\mathcal{F}^{\rm Fl}$, defined by Eq.~(\ref{eq:Flory}), remains invariant. The Born modulus [Eq.~(\ref{eq:Born})] hence reduces to $G^{\rm Born}\equiv\frac{1}{A}\,{\partial^2\mathring{\mathcal{F}}^{\rm ch}}/{\partial\gamma^2}|_{\gamma\to0}$. It is computed by writing  the shear stress at arbitrary $\gamma$:
\begin{equation}
\sigma_{xy}(\gamma)\equiv\frac{1}{A}\,\frac{\partial\mathring{\mathcal{F}}^{\rm ch}}{\partial\gamma}
= \frac{1}{A}\,\sum_{i<j} F_{ij}'(r_{ij})\frac{x_{ij}\,y_{ij}^0}{r_{ij}}\\
\end{equation}
and derivating once more. It will be useful to decompose the result as follows:
\begin{equation}\label{eq:GCs}
G^{\rm Born}=C+\sigma_{yy}^{\rm ch}
\end{equation}
with
\begin{equation}\label{eq:C}
C
=\frac{1}{A}\,\sum_{i<j} \left(F_{ij}''(r_{ij}^0)-\frac{F_{ij}'(r_{ij}^0)}{r_{ij}^0}\right)\left(\frac{x_{ij}^0y_{ij}^0}{r_{ij}^0}\right)^2\\
\end{equation}
and where
\begin{equation}
\sigma_{yy}^{\rm ch}=\frac{1}{A}\,\sum_{i<j} F_{ij}'(r_{ij}^0) \frac{(y_{ij}^0)^2}{r_{ij}^0}
\end{equation}
is the $yy$ component of the stress carried by the chain network in the initial, undeformed, state.

Let us note that the chain free-energy ${\mathcal{F}}^{\rm ch}$, which determines $G^{\rm Born}$, only depends on the distances $r_{ij}$ between connected nodes.
In such a case, the free-energy under deformation can be written as a function of the Green-Saint-Venant tensor $\tensor\eta=\frac{1}{2}(\tensor\Lambda^T\cdot\tensor\Lambda-\tensor 1)$, since $\vec r_{ij}^2-(\vec r_{ij}^0)^2=2\vec r_{ij}^0\cdot\tensor\eta\cdot\vec r_{ij}^0$, and the general elastic theory for discrete systems shows that the elastic stiffness tensor can be written as~\cite{LemaitreMaloney2006} $S_{\alpha\beta\kappa\chi}=C_{\alpha\beta\kappa\chi}+\sigma_{\beta\chi}\delta_{\alpha\kappa}$, where $C_{\alpha\beta\kappa\chi}=\frac{1}{A}{\partial^2\mathring{\mathcal{F}}}/{\partial\eta_{\alpha\beta}\partial\eta_{\kappa\chi}}$ is called the tensor of elastic constants, and $\tensor\sigma$ is the stress in the undeformed system. Expression~(\ref{eq:GCs}) corresponds exactly to this decomposition since $C$ is precisely the elastic constant $C_{xyxy}$. It should be emphasized that, because the Flory free-energy is invariant under simple shear and, consequently, $G^{\rm Born}$ determined by the variations of ${\mathcal{F}}^{\rm ch}$, the stress term is only $\sigma_{yy}^{\rm ch}$ but not the total $yy$ stress.

Finally, combining Eq.~(\ref{eq:S}) and (\ref{eq:GCs}) the total shear modulus reads:
\begin{equation}
G\equiv S_{xyxy}=\sigma_{yy}^{\rm ch}+C+G^{\rm NA}
\end{equation}
which leads to Eq.~(\ref{eq:G:decomp}) when the stress tensor is isotropic, a condition which, as we will see shortly, is satisfied by our triangular networks.
The non-affine contribution [Eq.~(\ref{eq:na})] involves the field of virtual forces $\{\vec\Xi_{i,xy}\}$ and the Hessian matrix $\mathcal{H}$. The explicit expression of $\mathcal{H}$ is provided in Appendix~\ref{app:Hessian}. Concerning $\{\vec\Xi_{i,xy}\}$, we note that, in view of Eq.~(\ref{eq:Xi}) and since the Flory free-energy is invariant under affine simple shear ($\partial{\mathring{\mathcal{F}}^{\rm Fl}}/{\partial\Lambda_{xy}}=0$), there is no Flory contribution to it.

It is worth noting that, since the Flory contributions to both $G^{\rm Born}$ and $\{\vec\Xi_{i,xy}\}$ vanish, excluded volume effects impact the linear elastic response only indirectly via the role they play in defining the equilibrium structure.

\section{A simple case: the homogeneous network}

We focus in this section on the case when all strands have an equal number of monomers $N$. Then, the nodes lie on a regular lattice at any level of deformation and non-affine effects are absent.

\subsubsection{General expressions}
The stress tensor $\tensor\sigma=\tensor\sigma^{\rm ch}+\tensor\sigma^{\rm Fl}$, in the undeformed state, is computed straighforwardly using~(\ref{eq:sigma:ch}) and~(\ref{eq:sigma:Fl}):
\begin{equation}\label{eq:homogeneous:ch}
\tensor\sigma^{\rm ch}=\sqrt{3}\,\frac{{F^{\rm ch}_N}'(\xi)}{\xi}\ \tensor1\equiv -P^{\rm ch}\,\tensor 1
\end{equation}
and
\begin{equation}\label{eq:pressure:flory}
\tensor\sigma^{\rm Fl}=-6k_BT(1-2\chi)\frac{a^2N^2}{\xi^4}\ \tensor1\equiv -P^{\rm Fl}\,\tensor 1
\end{equation}
As expected for a 2D triangular lattice, $\tensor\sigma^{\rm ch}$ is isotropic: $\sigma_{xx}^{\rm ch}=\sigma_{yy}^{\rm ch}=-P^{\rm ch}$. Note that changing the swelling state of our gel network amounts to varying $\xi$ at fixed $N$. Accordingly $P=P^{\rm ch}+P^{\rm Fl}=-{\partial\mathcal{F}}/{\partial A}$ is the osmotic pressure.

From equation~(\ref{eq:GCs}) we obtain for the shear modulus (which reduces to its Born estimate):
\begin{equation}
G=C-P^{\rm ch}
\end{equation}
with
\begin{equation}\label{eq:homogeneous:C}
C=\frac{\sqrt{3}}{4}\left({F^{\rm ch}_N}''(\xi)-\frac{{F^{\rm ch}_N}'(\xi)}{\xi}\right)
\end{equation}


\subsubsection{The Gaussian chain model}
Expressions~(\ref{eq:homogeneous:ch})-(\ref{eq:homogeneous:C}) take especially simple forms for the Gaussian model ($F^{\rm ch}=F^{\rm G}$).
We then get for the osmotic pressure:
\begin{equation}
P^{\rm G}=-\frac{\sqrt{3}k_BT}{Na\,\ell_p}\left(1-\left(\frac{\xi_{\rm eq}^{G}}{\xi}\right)^4\right)
\end{equation}
where $\xi_{\rm eq}^{\rm G}$, the mesh size at swelling equilibrium ($P=0$), reads:
\begin{equation}
\xi_{\rm eq}^{\rm G}/a=\left(2\sqrt{3}\ (1-2\chi)\frac{\ell_p}{a}\right)^{1/4}\!N^{3/4}
\end{equation}
It is seen to be proportional to $N^{3/4}$, the standard 2D scaling behavior expected in the Flory framework.
When $\xi<\xi_{\rm eq}^{\rm G}$, then $P<0$ and the gel is underswollen.

Turning to the value of the shear modulus, we first note that, since $F_N^{\rm G}$ is quadratic, the elastic constant $C^{\rm G}$ [Eq.~(\ref{eq:homogeneous:C})] vanishes so that:
\begin{equation}\label{eq:G:G}
G^{\rm G}=-P^{\rm ch, G}=\frac{\sqrt{3}k_BT}{Na\,\ell_p}
\end{equation}
We recover here the classical expression, deduced from rubber elasticity~\cite{RubinsteinColby2003}, which states that the shear modulus of a polymer gel is proportional to the chain number density times the elastic free-energy per chain.

\begin{figure}
\includegraphics[width=0.45\textwidth]{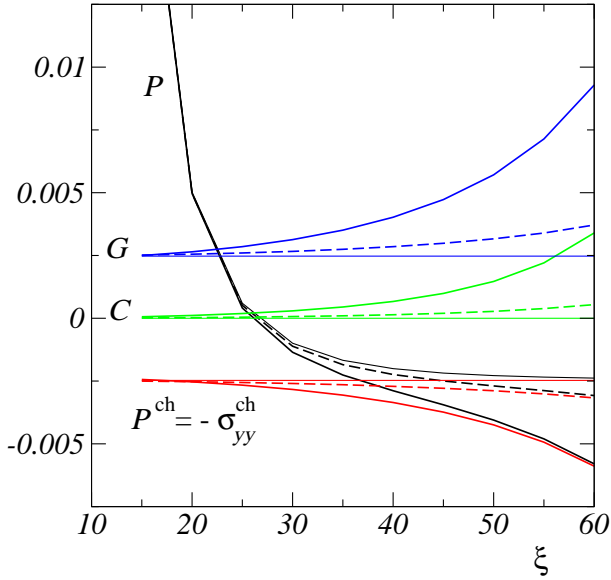}
\caption{\label{fig:homogeneous}
(Color online) Osmotic pressure $P$ (black), elastic constant $C$ (green), chain pressure $P^{\rm ch}$ (red) and shear modulus $G=C-P^{\rm ch}$ [Eq.~(\ref{eq:GCs})] (blue) plotted vs mesh size $\xi$,
for a homogeneous network of fixed structure ($N=100$) and the three models of chain free-energy: Gaussian (thin lines), FJ (dashed lines), and BT (thick solid lines). Parameters are:
$\ell_p=7a$, $1-2\chi=2.10^{-2}$.
The monomer size $a$ and $k_BT$ are taken as units of length and energy.
}
\end{figure}

\subsubsection{Dependence on swelling level for the three chain models}

Note that varying $\xi$ at fixed $N$ amounts to changing the swelling level of a gel of \emph{fixed network structure}.
So, the above expression means that, for the Gaussian chain model, $G$ is independent of the swelling level.

To compare the different chain models, we study a homogeneous network of fixed $N$, and plot on Fig.~(\ref{fig:homogeneous}) the four quantities $P$, $C$, $P^{\rm ch}$, and $G$ which characterize the gel mechanical state, versus the mesh size $\xi$, which characterizes the swelling level. The sharp drop of the osmotic pressure $P$ with swelling at small $\xi$ is essentially due to the decay of the Flory pressure $P^{\rm Fl}$ [Eq.~(\ref{eq:pressure:flory})], which is identical for all models.

Let us recall that the FJ and BT chain free-energies account for the stretch hardening of the chain segments while matching the Gaussian expression at small $\xi$. Expectedly, the values of $P$, $C$, $P^{\rm ch}$, and $G$ obtained with both models smoothly grow away from the Gaussian ones with increasing $\xi$. The BT model data exhibit a much steeper dependence on the swelling level, in agreement with the fact that it interpolates at high stretch with the WLC behavior which is stiffer than the FJ one.

The conditions of our study should be contrasted with the numerous existing works on filamentous networks\cite{Kroy2006,BroederszMacKintosh2014,PritchardHuangTerentjev2014}, which usually deal with the very stiff regime, $\ell_p\gg\xi\simeq Na$. Here, on the contrary, we consider much more flexible gels with the persistence length $\ell_p$ much smaller than the contour length $Na$, at moderate stretch ratio $\xi/(Na)$. It is thus striking that both the FJ and BT models lead to a very substantial growth of the shear modulus with swelling level, so that it departs from the Gaussian prediction by amplification factors $G/G^{\rm G}$ of respectively 1.5 (FJ) and 4.5 (BT) at our highest stretch ratio $\xi/(Na)=0.6$.

\begin{figure}
\includegraphics[width=0.45\textwidth]{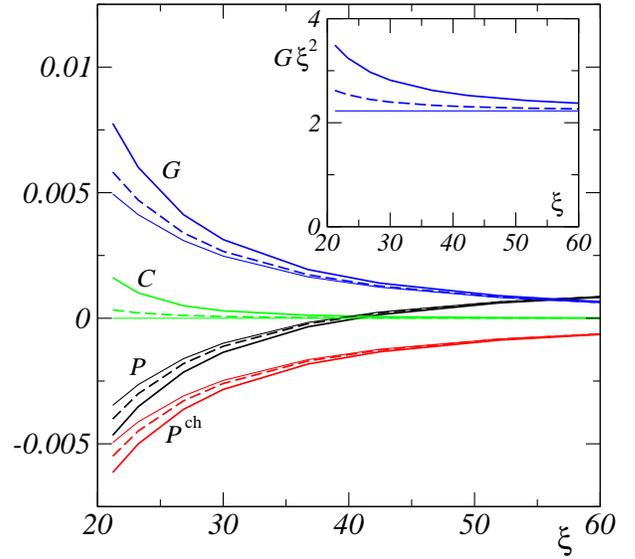}
\caption{\label{fig:homogeneous:scaling}
(Color online) Osmotic pressure $P$ (black), Born elastic constant $C$ (green), chain pressure $P^{\rm ch}$ (red) and shear modulus $G=C-P^{\rm ch}$ [Eq.~(\ref{eq:GCs})] (blue) plotted vs mesh size $\xi$,
for a homogeneous network at fixed monomer density ($2\sqrt{3}\,N/\xi^2\simeq0.38/a^2$) for the three models of chain free-energy: Gaussian (thin lines), FJ (dashed lines), and BT (thick solid lines). Parameters are:
$\ell_p=7a$, $1-2\chi=2.10^{-2}$. Insert: scaling plot, $G\,\xi^2$ vs $\xi$.
}
\end{figure}

\subsubsection{Dependence on cross-link density at fixed monomer concentration}

The progress of the crosslinking reaction under undrained conditions corresponds in our model to decreasing values of $\xi$ at fixed monomer density $\rho_m=2\,\sqrt{3}N/\xi^2$. Rubber elasticity theory~\cite{Treloar1949} then predicts that $G\sim1/\xi^d$, with $d$ the space dimension, an expression which is commonly used to estimate mesh sizes from the tracking of the shear modulus in experiments, yet is derived under the assumptions that chains are Gaussian and deformations affine.


We expect our Gaussian chain homogeneous network model to satisfy this scaling relation since it uphelds both assumptions.
Indeed, rewriting expression~(\ref{eq:G:G}) in terms of the monomer density yields:
\begin{equation}\label{eq:G:G}
G^{\rm G}=-P^{\rm ch, G}=\frac{6k_BT}{a\ell_p\rho_m}\,\frac{1}{\xi^2}
\end{equation}

Since this scaling derives from purely entropic arguments, it should break down in our FJ and BT models, which take stretch hardening into account.
To probe the magnitude of the expected deviations, we plot on Fig.~\ref{fig:homogeneous:scaling} $P$, $C$, $P^{\rm ch}$, and $G$ as a function of $\xi$ for networks of a fixed monomer density corresponding to $N=100$, $\xi=30a$, $\rho_m\simeq0.38/a^2$. As the crosslink density increases (decreasing $\xi$), the stretching ratio $\xi/(Na)=2\sqrt{3}/(\rho_ma\xi)$ grows, and so does the chain tension: $P^{\rm ch}$ is increasingly negative. Since $P^{\rm Fl}=P-P^{\rm ch}$ only depends on $\rho_m$, hence is $\xi$- and model-independent, we recover the intuitive result that osmotic pressure decreases together with $\xi$, i.e. that the swelling level grows with crosslinking.

Chain pressure $P^{\rm ch}$ is found to be only weakly model-dependent, i.e. is hardly affected by stretch hardening. As for the elastic constant $C$, its FJ value weakly departs from the Gaussian $C=0$ limit, but grows significantly at low $\xi$ in the BT model, in agreement with the fact that it specifically captures the deviations from harmonicity of the chain potential [see Eq.~(\ref{eq:homogeneous:C})]. The resulting departure of the shear modulus from the rubber elasticity scaling is quantified by plotting $G\,\xi^2$ vs $\xi$ in the insert of Fig.~\ref{fig:homogeneous:scaling}. Over the whole considered $\xi$ range (from $18\nm$ to $6\nm$) we find that $G^{\rm BT}/G^{\rm G}$ increases from 1.2 to 1.6, which correspond to quite substantial relative deviations.

\section{Disordered networks}

We now introduce disorder by attributing, as explained in~\ref{sec:disorder}, random values to the bond monomer numbers $N_{ij}$.
To isolate the effect of disorder strength, we vary the distribution width $\Delta$ for fixed average monomer and crosslink densities, i.e. fixed average $N=100$ and $\xi=30a$. The results, averaged over 100 configurations, are shown, for the bimodal distribution $B_{N,\Delta}$ [Eq.~(\ref{eq:bim})], on Fig.~\ref{fig:bim}.

\subsubsection{Osmotic and chain pressures}

Panel~(a) displays the osmotic pressure $P$ and its two terms, the chain and Flory pressures, $P^{\rm ch}$ and $P^{\rm Fl}$. Over the considered $\Delta$ range, $P^{\rm Fl}$ is not strictly constant, but its variations are much too small to be visible on the graph. The changes of osmotic pressure hence only result from the changes of $P^{\rm ch}$.

Visibly, $P^{\rm ch}$ decreases, i.e. the average chain tension increases with disorder. A hint to understand this trend is provided by constructing a \emph{regular network approximation} (RNA) which neglects the fact that the mechanically equilibrated network is distorted with respect to the reference triangular lattice. In this approximation, all end-to-end vectors assume the same length $\xi$ and orientations as in the homogeneous problem, and the chain pressure reduces to an expression similar to Eq.~(\ref{eq:homogeneous:ch}): $P^{\rm ch}_{\rm RNA}=-\sqrt{3}\,\langle{F_{N_{ij}}^{\rm ch}}'(\xi)\rangle/\xi$ where the average is taken over the $N_{ij}$ distribution. One easily checks that, for the Gaussian and FJ models, the chain tension $-{F_{N_{ij}}^{\rm ch}}'(\xi)$ is a concave function of $N_{ij}$ at all stretch ratios ($x<1$). The same holds for BT as long as $\ell_p<0.60\xi$, a condition satisfied by gels of flexible polymers at not too large underswelling levels. This property entails that (i) $P^{\rm ch}_{\rm RNA}<-\sqrt{3}\,{F_{N}^{\rm ch}}'(\xi)/\xi$, the chain pressure of the truly homogeneous network (for which $\Delta=0$, i.e. $N_{ij}=N$); (ii) at fixed $N$, $P^{\rm ch}_{\rm RNA}$ decreases (in algebraic value) with $\Delta$.

\begin{figure}
\includegraphics[width=0.45\textwidth]{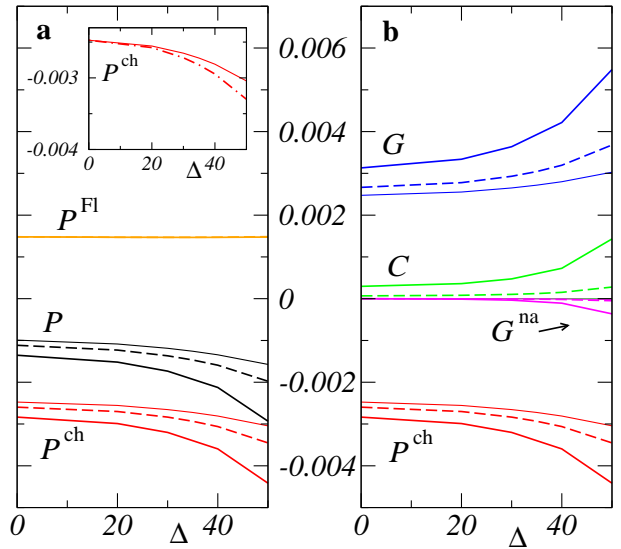}
\caption{\label{fig:bim}
(Color online) Network properties vs disorder strength, using bimodal monomer number distributions $B_{N,\Delta}$, with $N=100$ and varying $\Delta$. Parameters are: $\ell_p=7a$, $\xi=30a$, $1-2\chi=2.10^{-2}$. (a): decomposition of the osmotic pressure $P=P^{\rm Fl}+P^{\rm ch}$ (black) into its Flory (orange) and chain (red) contributions. Insert: $P^{\rm ch}$ (solid) vs $P^{\rm ch}_{\rm RNA}$ (dot-dashed) for the Gaussian model.
(b): the shear modulus $G=-P^{\rm ch}+C+G^{\rm na}$ (blue) and its three components [see Eq.~(\ref{eq:G:decomp})] $P^{\rm ch}$ (red), $C$ (green), and $G^{\rm na}$ (magenta). The three types of lines correspond to the Gaussian (thin solid), FJ (dashed), and BT (thick solid) models.
}
\end{figure}

We test this approximation in the simplest case of the Gaussian model by plotting $P^{\rm ch}_{\rm RNA}$ and the true $P^{\rm ch}$ of the inhomogeneous problem (insert of Fig.~\ref{fig:bim}{\bf a}): it appears that $P^{\rm ch}_{\rm RNA}$ noticeably overestimates the effect of disorder. To understand the origin of this discrepancy, let us write the microscopic expression of the chain pressure:
\begin{equation}
P^{\rm ch}=-\frac{1}{2A}\,\sum_{i<j} F_{ij}'(r_{ij}^0) r_{ij}^0=-\frac{\sqrt{3}}{\xi^2}\,\left\langle F_{ij}'(r_{ij}^0) r_{ij}^0\right\rangle
\end{equation}
where $\langle\cdot\rangle$ stands for the pair average. In the case of Gaussian chains, $P^{\rm ch}=-\frac{\sqrt{3}k_BT}{\xi^2\,a\ell_p}\,\langle r_{ij}^2/N_{ij}\rangle$, while $P^{\rm ch}_{\rm RNA}=-\frac{\sqrt{3}k_BT}{a\ell_p}\,\langle1/N_{ij}\rangle$. The comparison between $P^{\rm ch}$ and $P^{\rm ch}_{\rm RNA}$ points to the importance of correlations between chain end-to-end distances $r_{ij}$ and monomer numbers $N_{ij}$. Indeed, neglecting these correlations would lead to: $P^{\rm ch}=\langle r_{ij}^2\rangle/\xi^2\times P^{\rm ch}_{\rm RNA}$. Since, in a triangular network, $\langle r_{ij}^2\rangle>\xi^2$ always holds\footnote{
To check this inequality it suffices to notice that the energy of a network of Gaussian chains of unique $N$ is minimized, at fixed monomer density, i.e. fixed $\xi$, by the homogeneous state with $r_{ij}=\xi$.}, this assumption would result in $|P^{\rm ch}_{\rm RNA}|<|P^{\rm ch}|$, in contradiction with the data.

\begin{figure}
\includegraphics[width=0.45\textwidth]{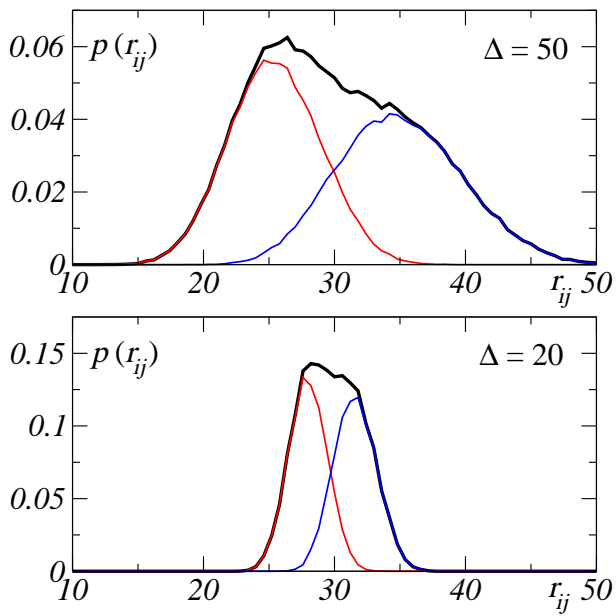}
\caption{\label{fig:bim:G}
(Color online) Black line: normalized distribution  of end-to-end distances $r_{ij}$ for an ensemble of disordered networks of Gaussian chains with the bimodal monomer number distribution $B_{100,50}$ (top) and $B_{100,20}$ (bottom). Red (blue): contribution of short (long) chains. Parameters: $\ell_p=7a$, $\xi=30a$, $1-2\chi=2.10^{-2}$.
}
\end{figure}
To evidence the correlations between chain end-to-end distances and monomer numbers, we plot on Fig.~\ref{fig:bim:G}, for two bimodal disorder strengths, the distribution of $r_{ij}$'s and its decomposition into the contributions of short and long chains. The two sub-distributions are clearly separated. Hence, in both cases, corresponding to large and modest elastic contrasts, chain lengths and monomer numbers are strongly correlated.

To understand why long (softer) chains are more extended than short (stiffer) ones, note that in the regular network approximation, the stretch ratios $s_{ij}=r_{ij}/(N_{ij}a)$ of long and short chains take the values $\xi/(a(N\pm\Delta))$: short chains are more taut (than long ones) and hence pull more strongly on their surroundings.
Under the effect of mechanical equilibration, short chains thus tend to contract while the long chains expand, which reduces the contrast of stretch ratios (and hence of chain tensions). Nevertheless, this effect does not fully resorb the stretch ratio contrast: the data of Fig.~\ref{fig:bim:G} show that, in mechanical equilibrium, the stretch ratio of the short chains peaks around 0.35 for $\Delta=20$ and 0.5 for $\Delta=50$, while those of the long chains peak around 0.26 for $\Delta=20$ and 0.23 for $\Delta=50$. In mechanical equilibrium, short chains always remain on average more taut than long ones, and increasingly so for large disorder strength. Mechanical equilibration hence only mitigates the effect of chain stiffness disorder as captured by the regular network approximation.

This observation sharply constrasts with the assumptions of rubber elasticity, which overlooks mechanical equilibration, and postulates that chains are independent, with a Gaussian statistics for end-to-end vectors. It follows that $\langle r_{ij}^2/N_{ij}\rangle$ is independent of $N_{ij}$, which leads to predicting a constant $P^{\rm ch}$. Since $C$ [see Eq.~(\ref{eq:C})] vanishes for Gaussian chains, rubber elasticity predicts the shear modulus to be disorder-independent.

\subsubsection{Contributions to shear modulus}

Let us now turn to panel~(b) of Fig.~\ref{fig:bim}, which displays the shear modulus along with the three terms of its decomposition $G=-P^{\rm ch}+C+G^{\rm na}$. In all cases, the chain pressure term $-P^{\rm ch}$ provides the largest contribution. The elastic constant $C$ captures departures from Gaussianity. It vanishes for Gaussian chains. It remains weak for the FJ model, is comparable to $|P^{\rm ch}|$ for the BT one, and in both cases grows with disorder strength. This latter effect results primarily from the fact that in both models $F''-F'/r$ is a concave function of $N$, a property which is expected to hold rather generally for stretch-hardening chain free-energies. Of course, like chain pressure, the exact value of $C$ in the FJ and BT models depends on the correlations evidenced in Fig.~\ref{fig:bim:G} between chain end-to-end distances and monomer numbers. Yet, as we saw above, mechanical equilibration does not destroy the constrast of stretch ratios between short (more taut) and long chains, and hence the value of $C$ grows with disorder strength.

The non-affine contribution $G^{\rm na}$ also increases with disorder, as intuitively expected. Yet, strikingly, it only contributes a small fraction of the shear modulus: $G^{\rm na}/G$ remains smaller than 1\% for both the Gaussian and FJ models; it reaches at most $\simeq 7\%$ for the highest disorder strength with the BT model. It thus turns out that the effect of disorder on the shear modulus is, for all practical purposes, captured by the Born approximation: $G\simeq-P^{\rm ch}+C$. As both terms grow similarly with disorder, so does the shear modulus $G$.

\subsubsection{Response to swelling}

\begin{figure}
\includegraphics[width=0.45\textwidth]{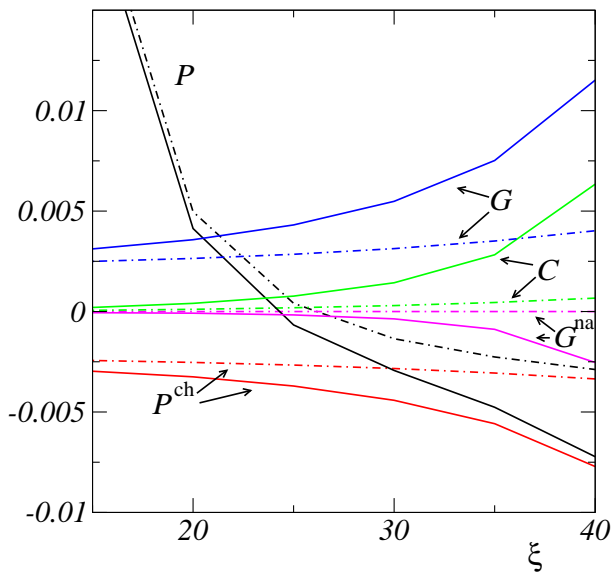}
\caption{\label{fig:bim:swelling}
(Color online) Dependence of network properties on the swelling level for the BT model, comparing the homogeneous $N=100$ network (dot-dashed) with a disordered network with $B_{100,50}$ (solid). Parameters: $\ell_p=7a$, $1-2\chi=2.10^{-2}$.
}
\end{figure}

In order to illustrate how disorder impacts the elastic response of the network to swelling, we compare on Fig.~\ref{fig:bim:swelling} the swelling response of a $B_{100,50}$ network with its homogeneous ($N=100$) counterpart, both using the BT model of chain free-energy. For pedagogical purposes, our swelling range extends up to a very high -- of course unrealistic -- level ($\xi=40$). All the measured quantities vary much more steeply with swelling level, in the disordered network. This can be ascribed to the hardening sensitivity of the short chains. For example, at $\xi=40$, the end-to-end distances of shorts chains are strongly peaked around $r_{ij}\simeq35$, which corresponds to a very high stretch ratio $\simeq0.7$. By contrast, the long chains exhibit rather modest stretch ratios $\lesssim0.4$, their end-to-end distances being broadly distributed about $r_{ij}\simeq50$. It is thus primarily the short chains that are responsible for the disordered-induced enhancement of stiffening with swelling.

\section{Conclusion}

Here we have analyzed in detail the case of flexible networks, in which the persistence length $\ell_p$ remains substantially smaller than the contour length of inter-crosslinks polymer chain segments. Their modulus can be decomposed into three terms, $G=-P^{\rm ch}+C+G^{\rm na}$
which present different sensitivities to chain flexibility and network disorder.
We have shown that:\\
(i) The non-affine contribution vanishes in the limit of either very flexible (Gaussian) chains or of homogeneous (non-disordered) networks. As a consequence, it contributes only a few percents of $G$ if we limit ourselves to a realistic range of disorder $(\Delta/N\lesssim 25\%)$ and swelling levels. Hence, the shear modulus is very well captured by the Born approximation: $G\simeq-P^{\rm ch}+C$\\
(ii) The elastic constant $C$ vanishes in the limit of flexible (Gaussian) chains, but is non-zero for stretch-hardening chains even in the absence of disorder. It thus presents values that are systematically larger that $G^{\rm na}$. However, it contributes only a small fraction of $G$ at reasonable disorder and swelling levels.\\
(iii) In all cases, $G$ is dominated by the chain pressure contribution $(-P^{\rm ch})$, all the more so that chains are more flexible.\\
This hierarchy of importance between the three contributions ($-P^{\rm ch}\gg C\gg G^{\rm na}$) is specific to flexible networks, in contrast with networks made of rigid polymers exhibiting finite average end-to-end distances on the scale of the mesh size $\xi$.

Of course, the specific value of $P^{\rm ch}$ (and of $G$) depends on both the distribution of chain stiffnesses and on the accomodation of this elastic disorder by mechanical equilibration. A trivial effect of disorder -- captured in the regular network approximation, which assumes that crosslinks lie on a regular lattice -- is that shorter (stiffer) chains tend to be more taut than long (softer) ones, which entails that disorder strongly enhances the stretch-hardening effects. Mechanical equilibration mitigates this effect by introducing correlations between end-to-end distances and chain contour lengths. These joint effects bring in a noticeable contribution to $G$ even in the case of fully-flexible (Gaussian) chains, in contradiction with the prediction of rubber elasticity.

\acknowledgements

We are grateful to Tristan Baumberger for stimulating our interest in the subject and for illuminating discussions.

\appendix

\section{Freely-jointed chain in 2D}
\label{app:fjc}

We compute the 2D polymer entropic elasticity following the usual route. A chain is supposed to comprise $N$ monomers of size $a$. To take into account the finite persistence length, it is decomposed into $N_K=N a/\ell_K$ Kuhn segments of fixed size $\ell_K=2\ell_p$ and orientation $\vec u_i$. The Kuhn segments are attached at their end points and can rotate and overlap freely. One end of the polymer is held fixed, and a force $\vec f= f\,\vec e_x$ is applied at the other, which amounts to introducing a potential $E=-\vec f\cdot\vec R$, with $\vec R=\ell_K\,\sum_i\vec u_i$ the end-to-end vector. Under equilibrium at temperature $T$, the partition function is $Z=z^{N_K}$ with
\begin{equation}
  \label{eq:z}
  z=\int_C\d\vec u_i\,e^{\beta \ell_K\,\vec f\cdot\vec u_i}=\int_{-\pi}^{\pi}\d\theta\,e^{\beta f\ell_K\,\cos\theta }=2\pi\,I_0(\beta f\ell_K)
\end{equation}
with $\beta=1/(k_BT)$ and $I_0$ a modified Bessel function of the first kind. The average end-to-end vector is:
\begin{equation}
  \label{eq:R}
  \left\langle\vec R\right\rangle=\frac{\partial \log Z}{\partial(\beta f)}\ \vec e_x=N_K \ell_K\,\frac{I_0'(\beta f\ell_K)}{I_0(\beta f\ell_K)}\ \vec e_x
\end{equation}
so that the norm of the end-to-end distance is $R=Nag(\beta f\ell_K)$, with $g(z)=I_0'(z)/I_0(z)$.

In order to invert approximately this relation, we note that for small $z$, at lowest order, $I_0(z)\simeq1+\frac{1}{4}\,z^2$, whence $g\simeq z/2$; and for large arguments, $I_0(z)\simeq e^z/\sqrt{2\pi\,z}$, whence $g\simeq1-1/(2z)$. The inverse relations, respectively $z\simeq2g$ (when $g\to0$) and $z\simeq\frac{1}{2(1-g)}$ (when $g\to1$). In the spirit of the Cohen approximation for the 3D case~\cite{Cohen1991}, we interpolate between these two limits using the rational function $z=g(1+\frac{1}{1-g^2})\equiv w'(g)$. So we write $f\simeq\frac{k_BT}{\ell_K}\,w'(R/(Na))$. From this, we derive the expression for the elastic the free-energy $F_{\rm FJ}$ of the freely jointed chain that appears in Eq.~(\ref{eq:fj}):
\begin{equation}
F^{\rm FJ}(R)=k_BT\,\frac{Na}{\ell_K}\,w\left(\frac{R}{Na}\right)
\end{equation}
where $\ell_K=2\ell_p$ and
\begin{equation}
w(x)=\frac{1}{2}\left(x^2-\log\left(1-x^2\right)\right)
\end{equation}

\section{Calculational details}
\label{app:delails}
The total free-energy is of the form:
$\mathcal{F}=\mathcal{F}^{\rm ch}+\mathcal{F}^{\rm Fl}$, with the chain and Flory contributions:
\begin{equation}\label{a:eq:fch}
\mathcal{F}^{\rm ch}=\sum_{\{ij\}} F_{ij}(r_{ij})
\end{equation}
and
\begin{equation}\label{a:eq:ffl}
\mathcal{F}^{\rm Fl}=\sum_{\{ijk\}} F_{ijk}(A_{ijk})
\end{equation}
In these expressions, $\{ij\}$ and $\{ijk\}$ respectively index pairs and triangles; $r_{ij}$ is the norm of the difference vector $\vec r_{ij}\equiv\vec r_j-\vec r_i$ between the positions of nodes $i$ and $j$; $A_{ijk}$ is the area of triangle $\{ijk\}$.
It is convenient, for any vector $\vec r=(x,y)$ to define $\vec r^\perp\equiv(-y,x)$; also, we assume -- without loss of generality -- that all triangles are oriented counterclockwise, so that the area of a triangle reads $A_{ijk}=\frac{1}{2}\,\vec r_{ij}^\perp\cdot\vec r_{ik}$.

\subsection{Microscopic forces}
The force exerted on node $i$ can be decomposed into the contributions of both the chain and Flory free-energies:
\begin{equation}
\vec f_i\equiv-\frac{\partial \mathcal{F}}{\partial\vec r_i}=\vec f_i^{\rm ch}+\vec f_i^{\rm Fl}
\end{equation}
The chain contribution $\vec f_i^{\rm ch}=-{\partial \mathcal{F}^{\rm ch}}/{\partial\vec r_i}$ is of the form:
\begin{equation}
\vec f_i^{\rm ch}=\sum_{\{jk\}}\ \vec f_{jk}\left(\delta_{ij}-\delta_{ik}\right)
\end{equation}
where the summation counts any pair once, and where $\vec f_{ij}$ denotes the force exerted by $j$ on $i$:
\begin{equation}
\vec f_{ij} \equiv -\frac{\partial F_{ij}(r_{ij})}{\partial \vec r_i}=
 {F_{ij}}'(r_{ij})\,\frac{\vec r_{ij}}{r_{ij}}=-\vec f_{ji}
\end{equation}
The last equation is Newton's second law.
The Flory contribution reads:
\begin{equation}
\vec f_i^{\rm Fl}=-\sum_{\{klm\}}\vec f_{klm\to i}
\end{equation}
where
\begin{equation}
\vec f_{klm\to i}=-\frac{\partial F_{klm}(A_{klm})}{\partial \vec r_i}
\end{equation}
is the force caused on node $i$, by the three body interaction between nodes $k$, $l$ and $m$. Of course, it is non-zero only if $i\in\{k,l,m\}$. Thanks to the translation and rotation invariance of $F_{klm}$, the forces it induces on the summits verify:
\begin{equation}\label{eq:sumf}
\vec f_{klm\to k}+\vec f_{klm\to l}+\vec f_{klm\to m}=0
\end{equation}
and
\begin{equation}
\vec r_k\times\vec f_{klm\to k}+\vec r_l\times\vec f_{klm\to l}+\vec r_m\times\vec f_{klm\to m}=0
\end{equation}
with $\times$ the vector product.
The general expression for $\vec f_{klm\to i}$ reads:
\begin{equation}
\vec f_{klm\to i}
=-F_{klm}'(A_{klm})\frac{\partial A_{klm}}{\partial\vec r_i}
\end{equation}
with
\begin{equation}\label{eq:aijkl}
\frac{\partial A_{klm}}{\partial\vec r_i}
= \frac{1}{2}\,\left(
\delta_{ik}\,\vec r_{lm}^\perp
+ \delta_{il}\,\vec r_{mk}^\perp
+ \delta_{im}\,\vec r_{kl}^\perp
\right)
\end{equation}
{}

\subsection{Macroscopic stress}

Let us consider an arbitrary deformation $\tensor\Lambda$. As in Section~\ref{sec:naff}, under the conditions of mechanical equilibrium, the nodes follow trajectories $\vec r_i(\tensor\Lambda)$ that are not identical in general to their affinely displaced values $\tensor\Lambda\cdot\vec r_i(\tensor 1)$; the non-affine displacements can be characterized by considering the zero strain antecedents: $\mathring{\vec r}_i(\tensor\Lambda)\equiv\tensor\Lambda^{-1}\cdot\vec r_i(\tensor\Lambda)$ during deformation.

The macroscopic Cauchy stress is defined as:
\begin{equation}
\tensor\sigma=\left.\frac{1}{A}\,\frac{\partial\mathcal{F}}{\partial\tensor\Lambda}\right|_{\tensor\Lambda\to\tensor 1}
=\left.\frac{1}{A}\,\left[
\frac{\partial\mathring{\mathcal{F}}}{\partial\tensor\Lambda}+
\frac{\partial\mathring{\mathcal{F}}}{\partial\mathring{\vec r}_i}\,\cdot\frac{\partial\mathring{\vec r}_i}{\partial\tensor\Lambda}
\right]\right|_{\tensor\Lambda\to\tensor 1}
\end{equation}
with $A$ the system area.
The second term vanishes in the rhs since ${\partial\mathring{\mathcal{F}}}/{\partial\mathring{\vec r}_i}$ does, by definition of mechanical equilibrium, so that:
\begin{equation}
\tensor\sigma=\left.\frac{1}{A}\,\frac{\partial\mathring{\mathcal{F}}}{\partial\tensor\Lambda}\right|_{\tensor\Lambda\to\tensor 1}\equiv
\tensor\sigma^{\rm ch}+
\tensor\sigma^{\rm Fl}
\end{equation}
This property, which is independent of the form of the total free-energy, states that the Cauchy stress can be computed by considering that deformation is associated with affine displacements only, since the derivative is taken at fixed $\mathring{\vec r}_i$.

The chain contribution is given by the Irvin-Kirkwood formula:
\begin{equation}
\tensor\sigma^{\rm ch}=\frac{1}{A}\,\sum_{\{ij\}}\vec f_{ij}\,\vec r_{ij}
\end{equation}
The Flory contribution is just:
\begin{equation}
    \tensor\sigma^{\rm Fl}
    =\frac{1}{A}\sum_{ijk}\,F'_{ijk}(A_{ijk})\,A_{ijk}\,\tensor1
\end{equation}

\subsection{Hessian}
The Hessian matrix
\begin{equation}
\mathcal{H}_{ij}=\left.\frac{\partial^2\mathring{\mathcal{F}}}{\partial\mathring{\vec r}_i\partial\mathring{\vec r}_j}\right|_{\tensor\Lambda=\tensor1}
\equiv\mathcal{H}_{ij}^{\rm ch}+\mathcal{H}_{ij}^{\rm Fl}
\end{equation}
is needed to compute the non-affine displacement fields from Eq.~(\ref{eq:naff}).
A concise expression for the pair (chain) contribution is obtain by writing how it applies on a displacement field ${\bf\vec u}\equiv\{\vec u_i\}$:
\begin{equation}
\left(\mathcal{H}^{\rm ch}\cdot{\bf\vec u}\right)_i\equiv\sum_j\mathcal{H}_{ij}^{\rm ch}\cdot\vec u_j
\end{equation}
The result is classical:
\begin{equation}\label{eq:hessian:ch}
\left(\mathcal{H}^{\rm ch}\cdot{\bf\vec u}\right)_i=-\sum_j\tensor M_{ij}\cdot\vec u_{ij}
\end{equation}
with $\vec u_{ij}=\vec u_j-\vec u_j$ and
\begin{equation}
\tensor M_{ij}=
\left(F_{ij}''-\frac{F_{ij}'}{r_{ij}}\right)\,\frac{\vec r_{ij}\,\vec r_{ij}}{r_{ij}^2}
+\frac{F_{ij}'}{r_{ij}}\,\tensor 1
\end{equation}

To compute the Flory contribution to the Hessian, we write:
\begin{equation}
\mathcal{H}_{ij}^{\rm Fl}=\sum_{\{klm\}} F_{klm}''\,\frac{\partial A_{klm}}{\partial\vec r_i}\,\frac{\partial A_{klm}}{\partial\vec r_j}
+F_{klm}'\,\frac{\partial^2 A_{klm}}{\partial\vec r_i\partial\vec r_j}
\end{equation}
and compute
\begin{equation}
\left(\mathcal{H}^{\rm Fl}\cdot{\bf\vec u}\right)_i=\sum_{j}\mathcal{H}_{ij}^{\rm Fl}\cdot\vec u_j
\end{equation}
The first gradient of $A_{klm}$ was provided under equation~(\ref{eq:aijkl}), and:
\begin{equation}
\begin{split}
\frac{\partial^2 A_{klm}}{\partial\vec r_i\partial\vec r_j} =
\frac{1}{2}\,\bigg(
&-\delta_{ik}\delta_{jl}+\delta_{il}\delta_{jk}\\
&-\delta_{im}\delta_{jk}+\delta_{ik}\delta_{jm}\\
&-\delta_{il}\delta_{jm}+\delta_{im}\delta_{jl}
\bigg)\,\tensor{S}
\end{split}
\end{equation}
where $\tensor S$ is the matrix with components $S_{\alpha\beta}\equiv\frac{\partial r^\perp_\alpha}{\partial r_\beta}$, i.e.:
\begin{equation}
\tensor S =
\left(
\begin{matrix}
0 & -1\\
1 & 0
\end{matrix}
\right)
\end{equation}
which transforms any vector $\vec a$ into $\tensor S\cdot\vec a=\vec a^\perp$.

Putting it all together, we find:
\begin{equation}\label{eq:hessian:Fl}
\begin{split}
\left(\mathcal{H}^{\rm Fl}\cdot{\bf\vec u}\right)_i&=\sum_j\mathcal{H}^{\rm Fl}_{ij}\cdot\vec u_j\\
&=\frac{1}{2}\sum_{\{klm\}}F''_{klm}\left(\vec r_{kl}^\perp\cdot\vec u_{lm}\!\!-\!\vec r_{lm}^\perp\cdot\vec u_{kl}\right)\frac{\partial A_{klm}}{\partial\vec r_i}\\
&+F'_{klm}\,\left(\delta_{ik}\vec u_{lm}^\perp+\delta_{il}\vec u_{mk}^\perp
+\delta_{im}\vec u_{kl}^\perp
\right)
\end{split}
\end{equation}

\subsection{The ${\bf\Xi}$ field}

To compute the fields $\{\vec\Xi_{i,\kappa\chi}\}$ defined by equation~(\ref{eq:Xi}), note that either equation~(\ref{eq:hessian:ch}) or~(\ref{eq:hessian:Fl}) the rhs is expressed entirely in terms of displacement differences (i.e. stretches) $\{\vec u_{ij}\}$, which is a field defined on pairs -- not on nodes. Thus the Hessian transform can be viewed a the series of two operations: a discrete gradient $\mathcal{D}: \{\vec u_i\}\to\{\vec u_{ij}=\vec u_j-\vec u_i\}$, followed by an operation denoted $\mathcal{M}$, which applies on the field of pair differences $\{\vec u_{ij}\}$.

This separation results from the fact that the free-energy function is only a function of the stretches $\{\vec u_{ij}\}$. The chain rule permits to write the Hessian as follows:
\begin{equation}
\mathcal{H}_{ij}=\left.\frac{\partial^2\mathring{\mathcal{F}}}{\partial\mathring{\vec r}_i\partial\mathring{\vec r}_j}\right|_{\tensor\Lambda=\tensor1}
=\sum_{\{kl\}}\left.\frac{\partial^2\mathring{\mathcal{F}}}{\partial\mathring{\vec r}_{i}\partial\mathring{\vec r}_{kl}}\cdot\frac{\partial\mathring{\vec r}_{kl}}{\partial\mathring{\vec r}_j}\right|_{\tensor\Lambda=\tensor1}
\end{equation}
Let us define
\begin{equation}
\tensor{\mathcal{M}}_{ikl}=\left.\frac{\partial^2\mathring{\mathcal{F}}}{\partial\mathring{\vec r}_{i}\partial\mathring{\vec r}_{kl}}\right|_{\tensor\Lambda=\tensor1}
\end{equation}
and
\begin{equation}
\mathcal{D}_{klj}=\left.\frac{\partial\mathring{\vec r}_{kl}}{\partial\mathring{\vec r}_j}\right|_{\tensor\Lambda=\tensor1}=\delta_{jl}-\delta_{jk}
\end{equation}
The transformation by $\mathcal{H}$ of any displacement field $\{\vec u_i\}$ can be written as: $\sum_j\mathcal{H}_{ij}\cdot\vec u_i=\sum_{\{kl\}} \mathcal{M}_{ikl}\cdot\sum_j \mathcal{D}_{klj}\vec u_j$, i.e. as the combination of two operators: $\mathcal{H}=\mathcal{M}\cdot\mathcal{D}$.
The discrete gradient $\mathcal{D}$ applies on any displacement field $\{\vec u_i\}$ and produces for each pair the value $\sum_j\mathcal{D}_{klj}\vec u_j=\vec u_l-\vec u_k$.

Operator $\mathcal{M}$ applies on ``stretch fields'', i.e. vector fields $\{\vec s_{kl}\}$ defined on all pairs, and transforms them into (force) fields carried by nodes with values: $(\mathcal{M}\cdot\{\vec s_{kl}\})_i\equiv\sum_{\{kl\}}\,\mathcal{M}_{ikl}\cdot\vec s_{kl}$. The explicit form of $\mathcal{M}=\mathcal{M}^{\rm ch}+\mathcal{M}^{\rm Fl}$ is easily obtained from the expressions derived above for $\mathcal{H}$. For an arbitrary field of stretches ${\bf\vec s}=\{\vec s_{ij}\}$, we find:
\begin{equation}
\left(\mathcal{M}^{\rm ch}\cdot{\bf\vec s}\right)_i=-\sum_j\tensor M_{ij}\cdot\vec s_{ij}
\end{equation}
and
\begin{equation}
\begin{split}
\left(\mathcal{M}^{\rm Fl}\cdot{\bf\vec s}\right)_i
&=\frac{1}{2}\sum_{\{klm\}}F''_{klm}\left(\vec r_{kl}^\perp\cdot\vec s_{lm}\!\!-\!\vec r_{lm}^\perp\cdot\vec s_{kl}\right)\frac{\partial A_{klm}}{\partial\vec r_i}\\
&+F'_{klm}\,\left(\delta_{ik}\vec s_{lm}^\perp+\delta_{il}\vec s_{mk}^\perp
+\delta_{im}\vec s_{kl}^\perp
\right)
\end{split}
\end{equation}
which compare with equations~(\ref{eq:hessian:ch}) or~(\ref{eq:hessian:Fl}).

With this in hand, let us turn back to the calculation of $\{\vec\Xi_{i,\kappa\chi}\}$. In Eq.~(\ref{eq:Xi}) the partial derivative with respect to strain refers to the changes in potential due to affine displacements, i.e. to variations of the difference vectors as ${\vec r}_{kl}^{\rm aff}(\tensor\Lambda)\equiv\tensor\Lambda\cdot{\vec r}_{kl}(\tensor1)$. Using the chain rule, we then find:
\begin{equation}
\vec\Xi_{i,\kappa\chi}=-\left.\frac{\partial^2\mathring{\mathcal{F}}}{\partial\Lambda_{\kappa\chi}\partial\mathring{\vec r}_i}\right|_{\tensor\Lambda\to\tensor1}=
-\sum_{\{kl\}}\mathcal{M}_{ikl}\left.\cdot\frac{\partial{\vec r}_{kl}^{\rm aff}}{\partial\Lambda_{\kappa\chi}}\right|_{\tensor\Lambda\to\tensor1}
\end{equation}
which is, up to the minus sign, the transform by $\mathcal{M}$ of a field of virtual stretches, $\left.\frac{\partial{\vec r}_{kl}^{\rm aff}}{\partial\Lambda_{\kappa\chi}}\right|_{\tensor\Lambda\to\tensor1}$, the value of which is easily computed on each pair.

%


\begin{thebibliography}{16}%
\makeatletter
\providecommand \@ifxundefined [1]{%
 \@ifx{#1\undefined}
}%
\providecommand \@ifnum [1]{%
 \ifnum #1\expandafter \@firstoftwo
 \else \expandafter \@secondoftwo
 \fi
}%
\providecommand \@ifx [1]{%
 \ifx #1\expandafter \@firstoftwo
 \else \expandafter \@secondoftwo
 \fi
}%
\providecommand \natexlab [1]{#1}%
\providecommand \enquote  [1]{``#1''}%
\providecommand \bibnamefont  [1]{#1}%
\providecommand \bibfnamefont [1]{#1}%
\providecommand \citenamefont [1]{#1}%
\providecommand \href@noop [0]{\@secondoftwo}%
\providecommand \href [0]{\begingroup \@sanitize@url \@href}%
\providecommand \@href[1]{\@@startlink{#1}\@@href}%
\providecommand \@@href[1]{\endgroup#1\@@endlink}%
\providecommand \@sanitize@url [0]{\catcode `\\12\catcode `\$12\catcode
  `\&12\catcode `\#12\catcode `\^12\catcode `\_12\catcode `\%12\relax}%
\providecommand \@@startlink[1]{}%
\providecommand \@@endlink[0]{}%
\providecommand \url  [0]{\begingroup\@sanitize@url \@url }%
\providecommand \@url [1]{\endgroup\@href {#1}{\urlprefix }}%
\providecommand \urlprefix  [0]{URL }%
\providecommand \Eprint [0]{\href }%
\providecommand \doibase [0]{http://dx.doi.org/}%
\providecommand \selectlanguage [0]{\@gobble}%
\providecommand \bibinfo  [0]{\@secondoftwo}%
\providecommand \bibfield  [0]{\@secondoftwo}%
\providecommand \translation [1]{[#1]}%
\providecommand \BibitemOpen [0]{}%
\providecommand \bibitemStop [0]{}%
\providecommand \bibitemNoStop [0]{.\EOS\space}%
\providecommand \EOS [0]{\spacefactor3000\relax}%
\providecommand \BibitemShut  [1]{\csname bibitem#1\endcsname}%
\let\auto@bib@innerbib\@empty
\bibitem [{\citenamefont {Kroy}(2006)}]{Kroy2006}%
  \BibitemOpen
  \bibfield  {author} {\bibinfo {author} {\bibfnamefont {K.}~\bibnamefont
  {Kroy}},\ }\href {\doibase 10.1016/j.cocis.2005.10.001} {\bibfield  {journal}
  {\bibinfo  {journal} {Curr. Opin. Colloid Interface Sci.}\ }\textbf {\bibinfo
  {volume} {11}},\ \bibinfo {pages} {56} (\bibinfo {year} {2006})}\BibitemShut
  {NoStop}%
\bibitem [{\citenamefont {Broedersz}\ and\ \citenamefont
  {MacKintosh}(2014)}]{BroederszMacKintosh2014}%
  \BibitemOpen
  \bibfield  {author} {\bibinfo {author} {\bibfnamefont {C.~P.}\ \bibnamefont
  {Broedersz}}\ and\ \bibinfo {author} {\bibfnamefont {F.~C.}\ \bibnamefont
  {MacKintosh}},\ }\href {\doibase 10.1103/RevModPhys.86.995} {\bibfield
  {journal} {\bibinfo  {journal} {Reviews of Modern Physics}\ }\textbf
  {\bibinfo {volume} {86}},\ \bibinfo {pages} {995} (\bibinfo {year}
  {2014})}\BibitemShut {NoStop}%
\bibitem [{\citenamefont {Pritchard}\ \emph {et~al.}(2014)\citenamefont
  {Pritchard}, \citenamefont {Huang},\ and\ \citenamefont
  {Terentjev}}]{PritchardHuangTerentjev2014}%
  \BibitemOpen
  \bibfield  {author} {\bibinfo {author} {\bibfnamefont {R.~H.}\ \bibnamefont
  {Pritchard}}, \bibinfo {author} {\bibfnamefont {Y.~Y.~S.}\ \bibnamefont
  {Huang}}, \ and\ \bibinfo {author} {\bibfnamefont {E.~M.}\ \bibnamefont
  {Terentjev}},\ }\href {\doibase 10.1039/c3sm52769g} {\bibfield  {journal}
  {\bibinfo  {journal} {Soft Matter}\ }\textbf {\bibinfo {volume} {10}},\
  \bibinfo {pages} {1864} (\bibinfo {year} {2014})}\BibitemShut {NoStop}%
\bibitem [{\citenamefont {Carrillo}\ \emph {et~al.}(2013)\citenamefont
  {Carrillo}, \citenamefont {MacKintosh},\ and\ \citenamefont
  {Dobrynin}}]{CarrilloMacKintoshDobrynin2013}%
  \BibitemOpen
  \bibfield  {author} {\bibinfo {author} {\bibfnamefont {J.~M.~Y.}\
  \bibnamefont {Carrillo}}, \bibinfo {author} {\bibfnamefont {F.~C.}\
  \bibnamefont {MacKintosh}}, \ and\ \bibinfo {author} {\bibfnamefont {A.~V.}\
  \bibnamefont {Dobrynin}},\ }\href {\doibase 10.1021/ma400478f} {\bibfield
  {journal} {\bibinfo  {journal} {Macromolecules}\ }\textbf {\bibinfo {volume}
  {46}},\ \bibinfo {pages} {3679} (\bibinfo {year} {2013})}\BibitemShut
  {NoStop}%
\bibitem [{\citenamefont {Meng}\ and\ \citenamefont
  {Terentjev}(2016)}]{MengTerentjev2016}%
  \BibitemOpen
  \bibfield  {author} {\bibinfo {author} {\bibfnamefont {F.~L.}\ \bibnamefont
  {Meng}}\ and\ \bibinfo {author} {\bibfnamefont {E.~M.}\ \bibnamefont
  {Terentjev}},\ }\href {\doibase 10.1039/c6sm01029f} {\bibfield  {journal}
  {\bibinfo  {journal} {Soft Matter}\ }\textbf {\bibinfo {volume} {12}},\
  \bibinfo {pages} {6749} (\bibinfo {year} {2016})}\BibitemShut {NoStop}%
\bibitem [{\citenamefont {{de Gennes}}(1979)}]{deGennes1979}%
  \BibitemOpen
  \bibfield  {author} {\bibinfo {author} {\bibfnamefont {P.-G.}\ \bibnamefont
  {{de Gennes}}},\ }\href@noop {} {\emph {\bibinfo {title} {{S}caling
  {C}oncepts in {P}olymer {P}hysics}}}\ (\bibinfo  {publisher} {Cornell
  University Press},\ \bibinfo {year} {1979})\BibitemShut {NoStop}%
\bibitem [{\citenamefont {Treloar}(1949)}]{Treloar1949}%
  \BibitemOpen
  \bibfield  {author} {\bibinfo {author} {\bibfnamefont {L.}~\bibnamefont
  {Treloar}},\ }\href@noop {} {\emph {\bibinfo {title} {{T}he {P}hysics of
  {R}ubber {E}lasticity}}}\ (\bibinfo  {publisher} {Clarendon Press},\ \bibinfo
  {year} {1949})\BibitemShut {NoStop}%
\bibitem [{\citenamefont {Onuki}(2002)}]{Onuki2002}%
  \BibitemOpen
  \bibfield  {author} {\bibinfo {author} {\bibfnamefont {A.}~\bibnamefont
  {Onuki}},\ }\href@noop {} {\emph {\bibinfo {title} {{P}hase {T}ransition
  {D}ynamics}}}\ (\bibinfo  {publisher} {Cambridge University Press},\ \bibinfo
  {year} {2002})\BibitemShut {NoStop}%
\bibitem [{\citenamefont {Rubinstein}\ and\ \citenamefont
  {Colby}(2003)}]{RubinsteinColby2003}%
  \BibitemOpen
  \bibfield  {author} {\bibinfo {author} {\bibfnamefont {M.}~\bibnamefont
  {Rubinstein}}\ and\ \bibinfo {author} {\bibfnamefont {R.}~\bibnamefont
  {Colby}},\ }\href@noop {} {\emph {\bibinfo {title} {{P}olymer {P}hysics}}}\
  (\bibinfo  {publisher} {Oxford University Press},\ \bibinfo {year}
  {2003})\BibitemShut {NoStop}%
\bibitem [{\citenamefont {Cohen}(1991)}]{Cohen1991}%
  \BibitemOpen
  \bibfield  {author} {\bibinfo {author} {\bibfnamefont {A.}~\bibnamefont
  {Cohen}},\ }\href {\doibase 10.1007/BF00366640} {\bibfield  {journal}
  {\bibinfo  {journal} {Rheologica Acta}\ }\textbf {\bibinfo {volume} {30}},\
  \bibinfo {pages} {270} (\bibinfo {year} {1991})}\BibitemShut {NoStop}%
\bibitem [{\citenamefont {Blundell}\ and\ \citenamefont
  {Terentjev}(2009)}]{BlundellTerentjev2009}%
  \BibitemOpen
  \bibfield  {author} {\bibinfo {author} {\bibfnamefont {J.~R.}\ \bibnamefont
  {Blundell}}\ and\ \bibinfo {author} {\bibfnamefont {E.~M.}\ \bibnamefont
  {Terentjev}},\ }\href {\doibase 10.1021/ma9004633} {\bibfield  {journal}
  {\bibinfo  {journal} {Macromolecules}\ }\textbf {\bibinfo {volume} {42}},\
  \bibinfo {pages} {5388} (\bibinfo {year} {2009})}\BibitemShut {NoStop}%
\bibitem [{Note1()}]{Note1}%
  \BibitemOpen
  \bibinfo {note} {In order to ensure that $F_{ij}^{\protect \rm BT}$ matches
  $F_{ij}^{\protect \rm G}$ in the $Na/\ell _p\gg 1$, $x\ll 1$ limit, our
  definition of the persistence length differs by a factor of $4/\pi $ from
  Blundell and Terentjev's}\BibitemShut {NoStop}%
\bibitem [{\citenamefont {Bohidar}\ and\ \citenamefont
  {Jena}(1994)}]{BohidarJena1994}%
  \BibitemOpen
  \bibfield  {author} {\bibinfo {author} {\bibfnamefont {H.~B.}\ \bibnamefont
  {Bohidar}}\ and\ \bibinfo {author} {\bibfnamefont {S.~S.}\ \bibnamefont
  {Jena}},\ }\href {\doibase 10.1063/1.467004} {\bibfield  {journal} {\bibinfo
  {journal} {Journal of Chemical Physics}\ }\textbf {\bibinfo {volume} {100}},\
  \bibinfo {pages} {6888} (\bibinfo {year} {1994})}\BibitemShut {NoStop}%
\bibitem [{\citenamefont {Lema\^{\i}tre}\ and\ \citenamefont
  {Maloney}(2006)}]{LemaitreMaloney2006}%
  \BibitemOpen
  \bibfield  {author} {\bibinfo {author} {\bibfnamefont {A.}~\bibnamefont
  {Lema\^{\i}tre}}\ and\ \bibinfo {author} {\bibfnamefont {C.}~\bibnamefont
  {Maloney}},\ }\href@noop {} {\bibfield  {journal} {\bibinfo  {journal} {J.
  Stat. Phys.}\ }\textbf {\bibinfo {volume} {123}},\ \bibinfo {pages} {415}
  (\bibinfo {year} {2006})}\BibitemShut {NoStop}%
\bibitem [{\citenamefont {Barron}\ and\ \citenamefont
  {Klein}(1965)}]{BarronKlein1965}%
  \BibitemOpen
  \bibfield  {author} {\bibinfo {author} {\bibfnamefont {T.}~\bibnamefont
  {Barron}}\ and\ \bibinfo {author} {\bibfnamefont {M.}~\bibnamefont {Klein}},\
  }\href@noop {} {\bibfield  {journal} {\bibinfo  {journal} {Proc. Phys. Soc.}\
  }\textbf {\bibinfo {volume} {85}},\ \bibinfo {pages} {523} (\bibinfo {year}
  {1965})}\BibitemShut {NoStop}%
\bibitem [{Note2()}]{Note2}%
  \BibitemOpen
  \bibinfo {note} {To check this inequality it suffices to notice that the
  energy of a network of Gaussian chains of unique $N$ is minimized, at fixed
  monomer density, i.e. fixed $\xi $, by the homogeneous state with $r_{ij}=\xi
  $.}\BibitemShut {Stop}%
\end{thebibliography}
\end{document}